\documentclass[conference]{IEEEtran}
\ifCLASSOPTIONcompsoc
 \usepackage[nocompress]{cite}
\else
 \usepackage{cite}
\fi
\ifCLASSINFOpdf
\else
\fi
\usepackage{color}
\usepackage{float}
\usepackage{graphicx}
\usepackage{subfig}
\usepackage{soul}
\usepackage{comment}
\graphicspath{{Pictures/}}
\begin{document}

%
\title{Detecting Anomalies using Overlapping Electrical Measurements in Smart Power Grids}

\author{\IEEEauthorblockN{Sina Sontowski\IEEEauthorrefmark{1}, Nigel Lawrence\IEEEauthorrefmark{2}, Deepjyoti Deka\IEEEauthorrefmark{3}, and Maanak Gupta\IEEEauthorrefmark{4}}
\IEEEauthorblockA{\IEEEauthorrefmark{1}\IEEEauthorrefmark{4}{Dept. of Computer Science},
{Tennessee Technological University},
Cookeville, Tennessee, USA \\\IEEEauthorrefmark{2}\IEEEauthorrefmark{3}Los Alamos National Laboratory, Los Alamos, New Mexico, USA\\}
\IEEEauthorrefmark{1}ssontowsk42@tntech.edu,
\IEEEauthorrefmark{2}nlawrence@lanl.gov, 
\IEEEauthorrefmark{3}deepjyoti@lanl.gov,
\IEEEauthorrefmark{4}mgupta@tntech.edu}


%


\maketitle

\begin{abstract}
As cyber-attacks against critical infrastructure become more frequent, it is increasingly important to be able to rapidly identify and respond to these threats. This work investigates two independent systems with overlapping electrical measurements with the goal to more rapidly identify anomalies. The independent systems include HIST, a SCADA historian, and ION, an automatic meter reading system (AMR). While prior research has explored the benefits of fusing measurements, the possibility of overlapping measurements from an existing electrical system has not been investigated. To that end, we explore the potential benefits of combining overlapping measurements both to improve the speed/accuracy of anomaly detection and to provide additional validation of the collected measurements. In this paper, we show that merging overlapping measurements provide a more holistic picture of the observed systems. By applying Dynamic Time Warping more anomalies were found -- specifically, an average of 349 times more anomalies, when considering anomalies from both overlapping measurements. When merging the overlapping measurements, a percent change of anomalies of up to 785\% can be achieved compared to a non-merge of the data as reflected by experimental results.
\end{abstract}


%
\IEEEpeerreviewmaketitle

\section{Introduction}
In recent years, cyber attacks in critical infrastructures have become more frequent \cite{sontowski2020cyber,hahn2011cyber,gupta2018authorization,le2020gridattacksim} and increased in complexity and sophistication. Smart power grids are vulnerable to these attacks in large part due to an increase in cyber-physical connectivity in Supervisory Control and Data Acquisition (SCADA) systems that are used to monitor and control parts of the power grid \cite{zhu2011taxonomy}. While newer digital devices, even at the edge of the grid such as micro-phasor measurement units ($\mu$-PMUs) \cite{zhou2016abnormal,deka2015optimal}, can increase connectivity, control, and estimation tasks used in power grid operation, they also increase the risk from potential cyber threats \cite{8762213}. An unobserved or hidden cyber attack on the power grid can de-energize power system components and aggregate operating conditions by causing overloading and instability \cite{6003810}. The 2015\footnote{https://www.wired.com/2016/03/inside-cunning-unprecedented-hack-ukraines-power-grid/} Ukraine power grid hack exemplifies the threat that cyber attacks can pose to the power grid. Many consumers in Ukraine temporarily lost power due to a cyber attack exploiting Windows vulnerability CVE-2014-4114. A phishing email was opened that installed BlackEnergy malware on the system. This was not a zero-day vulnerability and the operator should have been aware of the vulnerability in the system and taken action beforehand \cite{8762213}.
While anomaly detection solutions would have not prevented this attack, they would have alerted the operator about unusual activity in the system, potentially allowing them to respond more rapidly. Accordingly, anomaly detection identifies potential issues in the system and can therefore enable operators to respond to malicious activity as it happens \cite{chandola2009anomaly}. Thus, it allows for an early detection of cyber-intrusions, and can be understood as an early warning mechanism \cite{6003810}. 
In general, anomaly detection refers to the problem of finding patterns in data that do not conform to expected behavior, and it is being widely used in different domains \cite{moghaddass2017hierarchical,gupta2021detecting,fenza2019drift, cook2019anomaly,adkisson2021autoencoder,gupta2021hierarchical}. 

\begin{figure}[!t]
  \centering
  \includegraphics[width=\columnwidth]{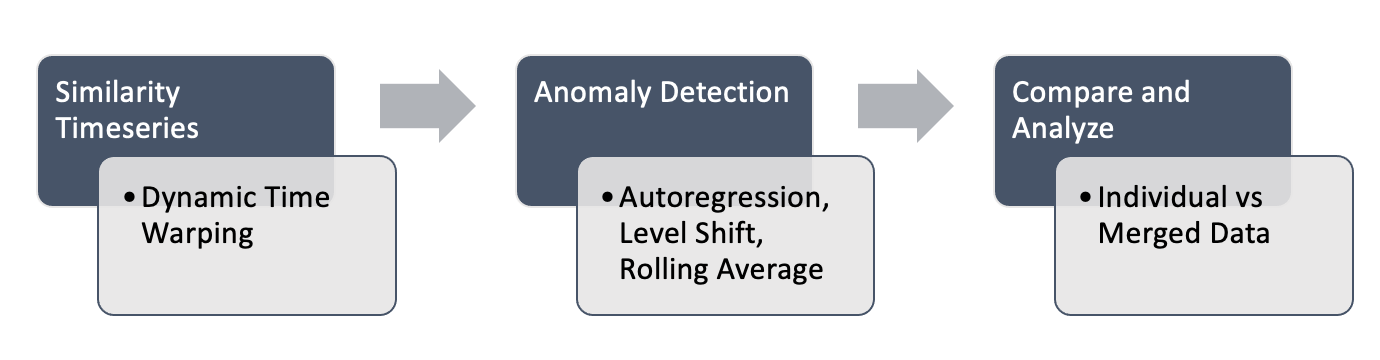}
  \caption{Overview of Proposed Approach}
  \label{fig:Approach}
\end{figure}

The objective of this paper is to develop improved techniques for anomaly detection in smart power grids using electrical measurements by which operators can be alerted early of any abnormal activity, and action can be taken to ameliorate from that state. 
Specifically, we present an approach to improve the speed and accuracy of anomaly detection by using overlapping electrical measurements from two independent systems taken at different points throughout the United States Department of Energy Los Alamos National Laboratory's power grid. Our approach consists of three parts as illustrated in Figure \ref{fig:Approach}. First, we determine the overlapping measurements by using Dynamic Time Warping (DTW). Then unsupervised anomaly detection algorithms including autoregression, level shift, and rolling average are applied to the overlapping measurements. This approach differs from existing work (as discussed in related work section) as the overlapping electrical measurements used for anomaly detection are from a real power grid. To the best of our knowledge, no existing work has applied DTW as a way to improve anomaly detection. In the final stage, we analyze the anomalies that were discovered through merging the overlapping measurements, and we compare them to their individual anomalies.

The main contributions of this paper are as follows:
\begin{itemize}
  \item We elaborate approaches used for dealing with overlapping electrical measurements, in addition to verifying them.
  \item  We compare results of different unsupervised anomaly detection algorithms applied on real-life electrical measurements.
  \item We conjecture that merging overlapping measurements for anomaly detection gives a more holistic view and finds more anomalies.
\end{itemize}

The rest of the paper has been organized as follows. In Section \ref{sec:related}, the paper explores related works and their significance. In Section \ref{sec:background}, background information and preliminaries are provided for a better understanding of our implementation described in Section \ref{sec:dtw} and \ref{sec:AD}. A conclusion is provided and possible future work approaches are mentioned in Section \ref{sec:conc} and Section \ref{sec:fw}, respectively. 

\section{Related Work}
\label{sec:related}
A neuro-inspired architecture called Hierarchical Temporal Memory (HTM) was developed to perform unsupervised anomaly detection \cite{9096053}. HTM learns sparse distributed temporal representation of sequential data which is shown to achieve competitive scores for real-time anomaly detection compared to state-of-the-art approaches. Anomaly detection is also performed on unlabeled electrical measurements ($\mu$-PMU) from the smart grid. HTM was able to capture spatial and temporal anomalies in the $\mu$-PMU data from the power grid. While electrical measurements ($\mu$-PMU data) were used for this approach, the anomaly detection is real-time, involves hierarchical temporal memory, but does not include overlapping electrical measurements as proposed in our work. 

Researchers have used dynamic time warping (DTW) related to the power grid; however, Elafoudi et al. \cite{6877810}, \cite{9299915} apply it to smart meter readings for lowering the complexity of power disaggregation. Ausmus et al. \cite{9299915} apply DTW to electric utility data to cluster the electric utility net data based on the distance measure to improve operation for the next day. None of these papers apply DTW with the goal to improve anomaly detection.
Another work by Diab et al. uses DTW for anomaly detection \cite{8919604}. Instead of using electrical measurements, it used network traffic data to detect network anomalies. The network traffic is decomposed into control and data planes. Based on the DTW distance between these two, the network activities are classified as either benign or anomalous. 
The goal of Zheng's et al.'s work \cite{9178465} is to detect road anomalies based on acceleration data. The data windows of various length are compared with DTW. Anomalies are automatically identified by machine learning algorithms, and the types are distinguished with DTW. Their method improved the time consumption of a random forest filter. Although, this work uses DTW and has a similar goal of improving the speed of anomaly detection, it uses road acceleration data and not electrical measurements.

Autoregressive, rolling average, and level shift processes are common anomaly detection approaches. In Zhou's and Li's work \cite{AutoregressionNetwork} multilevel autoregression is applied on network traffic for anomaly detection. This method proved to be successful by correctly detecting more than 95\% of network anomalies. Often, autoregressive processes and rolling average are combined as illustrated in \cite{ARIMACloud} and \cite{ARIMANetwork}. Schmidt et al. \cite{ARIMACloud} apply the ARIMA algorithm, which is a combination of autoregression and rolling average, on real-time cloud monitoring data. Yaacob et al. \cite{ARIMANetwork} use ARIMA for network anomaly detection. Solomentsev et al. \cite{ChangePointRadar} apply level shift detection during radar operation, while Geng and Lai \cite{ChangePointSensor} apply it on sensor networks.

In summary, the related work does not address the challenge of improving anomaly detection based on overlapping electrical measurements, the core idea of our proposed work.
\section{Preliminaries and Background}
\label{sec:background}
\subsection{Dataset}
\begin{figure}[!t]
  \centering
  \includegraphics[width=\columnwidth]{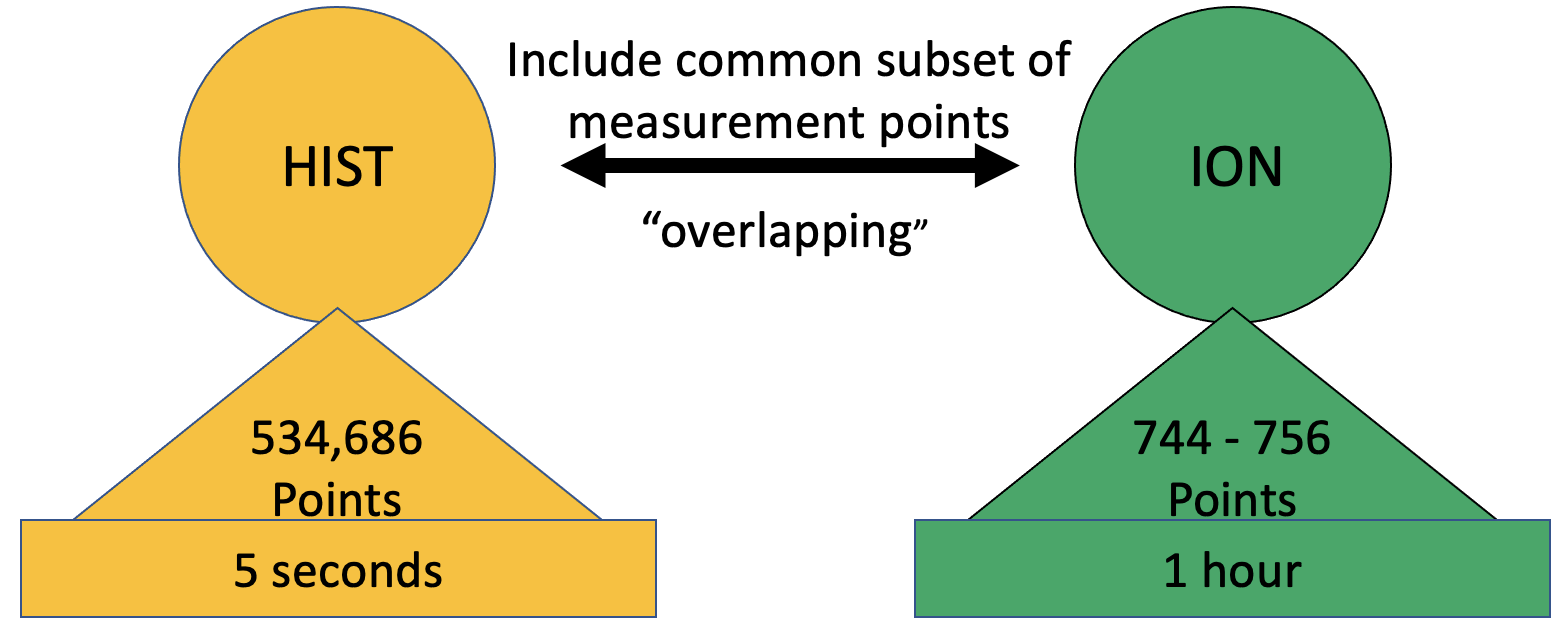}
  \caption{Overview of Dataset}
  \label{fig:Data}
\end{figure}

The data set used for this paper originates from Los Alamos National Laboratory's power grid from two independent overlapping electrical measurement systems: ION and HIST. HIST is a SCADA historian and records all of the data measured by the SCADA system. ION is an automatic meter reading (AMR) system that pulls electrical metering information from different meters and records them. As a result, the SCADA system does not measure all of the ION points, while the ION system does not measure all of the SCADA points. Therefore, they are two separate systems that share measurements for some meters, but not all of them. The data set was collected in December 2020 and contains 744-756 ION data points and 534,686 HIST data points for each time series. HIST data is collected at a higher frequency (approximately every 5 seconds) while ION is collected at a much lower frequency (approximately every hour). As illustrated in Figure \ref{fig:Data}, a subset of measurement points are common to both systems and are therefore said to be overlapping.

\subsection{Dynamic Time Warping}
The DTW algorithm is utilized to identify the overlapping measurements. It finds the similarity between two time series by calculating the distance between them. The lower the distance, the more similar the two time series are to each other. DTW involves a non-linear optimal alignment which ensures that similar time series match each other even if they are out of phase on the x-axis or require compression or expansion. This is different from Euclidean distance, where the i-th point in one time series is aligned with the i-th point in the other (one-to-one mapping). Instead, the DTW algorithm can make one-to-many mappings between points \cite{DTWMueen}. To find the best alignment, the DTW applies a warping function which minimizes the distance between the two time series. In our case, DTW is implemented to account for the difference in frequency between the two systems as well as any potential minor discrepancies in clock synchronization. DTW has applications in many domains such as robotics, data mining, and manufacturing \cite{salvador2004fastdtw}.

\begin{figure*}[!t]
\centering
\subfloat[First Ranked Match]{
 \includegraphics[width=85mm]{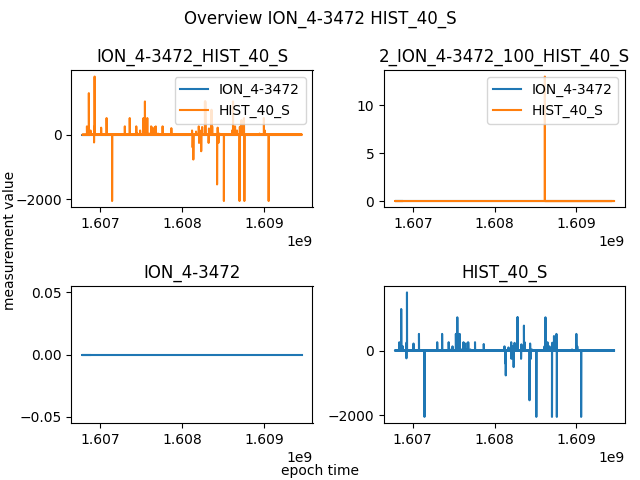}
}
\subfloat[Second Ranked Match]{
 \includegraphics[width=85mm]{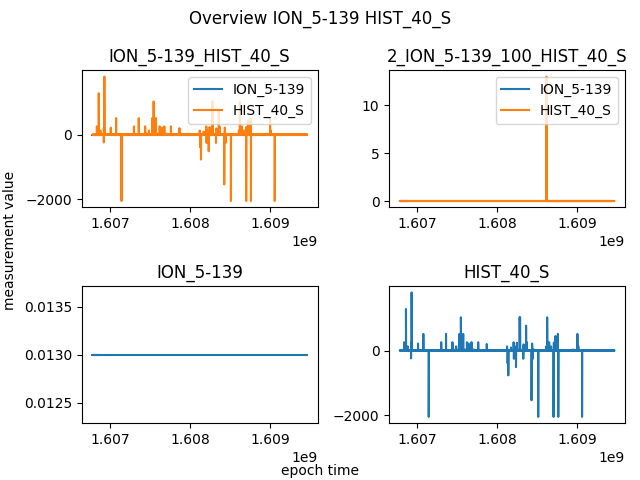}
}
\hspace{0mm}
\subfloat[Third Ranked Match]{
 \includegraphics[width=85mm]{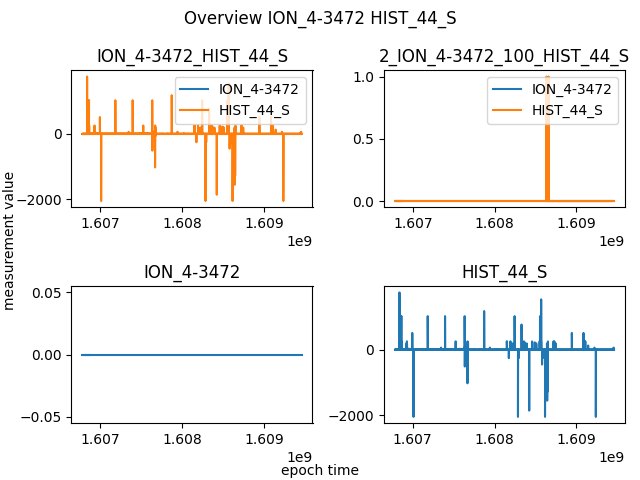}
}
\subfloat[Fourth Ranked Match]{
 \includegraphics[width=85mm]{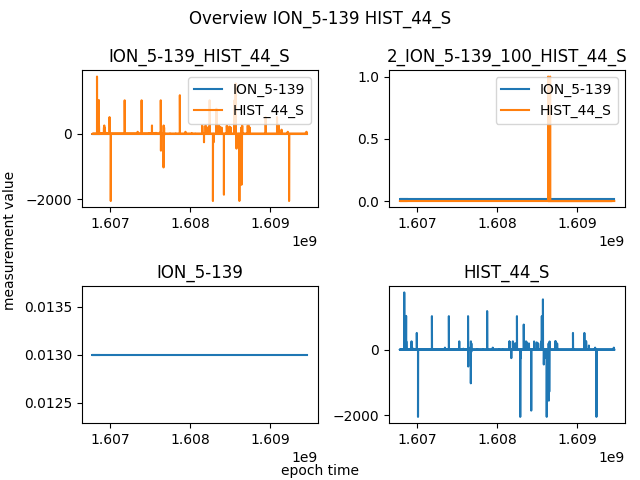}
}

\caption{Top Four Results Step-size.} 
Each ranked match (a), (b), (c), and (d) includes four graphs. The top-left graph includes both measurements drawn together, while the top-right graph draws their step-size sample. The bottom two graphs for each match show the actual measurements drawn independent from each other. 
\label{fig: 4 results}
\end{figure*}

\subsection{Unsupervised Anomaly Detection}
Unsupervised anomaly detection is executed on the overlapping electrical measurements. Anomaly detection, also called outlier detection, is applied in various fields to find irregularities within data. Anomaly detection approaches slightly differ depending on the data set, for example, a time series will have modified algorithms compared to a multidimensional data facet \cite{9166366}. The anomaly detection algorithms implemented in this paper to compare anomalies between the overlapping time series include autoregression (AR), level shift (LS), and rolling average (RA). All of these algorithms are statistical unsupervised anomaly detection approaches, which means they are used to find outliers within unlabeled time series data sets.
\subsubsection{Autoregression (AR)}
AR looks at autoregressive behavior changes to detect anomalies. AR models are used for time series analysis. Basically, linear regression is applied on the current data series and predictions for future values are created based on past values. Similar to linear regression, the outcome variable (Y) at a certain point in time is related to the predictor variable (X). However, for AR models past values of (Y) are factored into predicting (Y). Therefore, behavior is modeled based on past data \cite{medico2020,hannon2021real}. 
\subsubsection{Level Shift (LS)}
Level shift detects level shift outliers, which are often represented by a step function. These outliers occur when a sudden change in the mean level move an outlier and its following data onto a new level. This can be seasonal or not. Level shift is often used when the data has a lot of outliers, because it is not as sensitive to spikes in data. It works by taking two sliding windows and comparing their median values to detect a shift of values \cite{medico2020}.
\subsubsection{Rolling Average (RA)} Rolling average or Moving Average is similar to the AR model. However, instead of using past values of (Y) to predict (Y), past forecast errors are used to predict (Y) \cite{medico2020}.

\section{Dynamic Time Warping}
\label{sec:dtw}
\subsection{Implementation}
To determine similarity among the two different systems - HIST and ION, the DTW algorithm is used. Due to the overhead of the traditional DTW, we instead chose to use the fastDTW \cite{salvador2004fastdtw} implementation from the DTAIDistance library. Additionally, several sampling strategies were implemented to reduce run-time as opposed to using each time series in its entirety.

We considered three approaches to obtain a viable sample of the data:
\begin{enumerate}
 \item Step size, which includes collecting data points at different steps, and examples can be seen in Figure \ref{fig: 4 results}, illustrated in the top-right graph for each overall result in (a)-(d),
 \item Certain amount of points, for example, collecting the first 100 points, and
 \item Range of dates, for example, a day's worth of data.
\end{enumerate}

\subsubsection{Step Size}

\begin{table}[!t]
\begin{center}
\begin{tabular}{ |c|c|c|c| } 
 \hline
 Run Number & HIST step size & ION step size & Run-time (s) \\ 
 \hline
 1 & 100 & 2 & 2318.6 \\ 
 2 & 1000 & 1 & 616.7 \\ 
 3 & 1000 & 2 & 342.8 \\
 4 & 2000 & 2 & 222.6 \\
 5 & 3000 & 4 & 108.6 \\
 6 & 5000 & 7 & 65.7 \\
 \hline
\end{tabular}
\end{center}
\caption{Step Size Running Times.}
\label{tab:step-running-times}
\end{table}
For step-size, we collected a total of 6 runs, which are depicted in Table \ref{tab:step-running-times}. DTW can handle sequences of different lengths to the same ability as equal length sequences. In fact, there is no significant difference in accuracy between equal-length sequences and variable-length sequences \cite{ratanamahatana2005}. Therefore, the step-size was reduced as much as possible to include the most points without regard to the measured frequencies. However, the smaller the steps, the more data points, and the longer the DTW algorithm takes to compare all of the time series.

\subsubsection{Amount of Points}
For certain amount of points, we completed two runs total, where we took the first 100 and 200 points as illustrated in Table \ref{tab:amount-running-times}. Similar to the step runs, as the amount of points increase, the run-time increases, too.

\begin{table}[!t]
\begin{center}
\begin{tabular}{ |c|c|c| } 
 \hline
 Run Number & Point Amount & Run-time (s) \\ 
 \hline
 7 & 100 & 24.3 \\ 
 8 & 200 & 38.6 \\ 
 \hline
\end{tabular}
\end{center}
\caption{Point Amount Running Times.}
\label{tab:amount-running-times}
\end{table}

\begin{table}[!t]
\begin{center}
\begin{tabular}{ |p{1.2cm}|p{1.2cm}|p{1.2cm}|p{1.2cm}|p{1.2cm}| } 
 \hline
 Run Number & Date Range & HIST step size & ION step size & Run-time (s) \\ 
 \hline
 9 & 3 days & 50 & 1 & 191.8 \\ 
 \hline
\end{tabular}
\end{center}
\caption{Date Range Running Times.}
\label{tab:range-running-times}
\end{table}

\begin{figure}
\centering
\subfloat[step size 100 HIST, 2 ION]{
 \includegraphics[width=75mm]{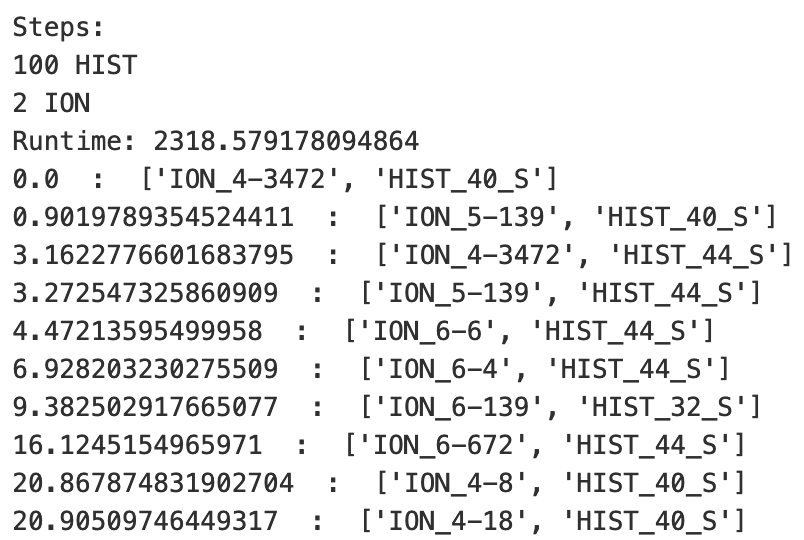}
}
\newline
\subfloat[three day range and step size]{
 \includegraphics[width=75mm]{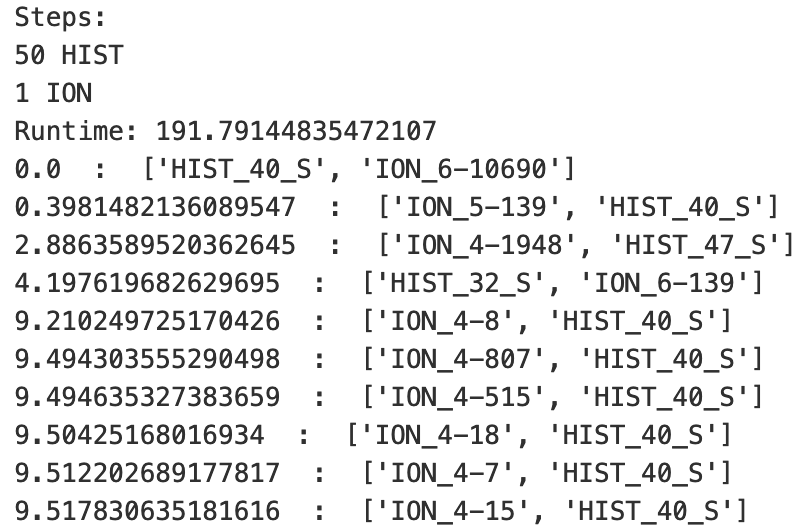}
}
\hspace{0mm}
\subfloat[first 200 points]{
 \includegraphics[width=75mm]{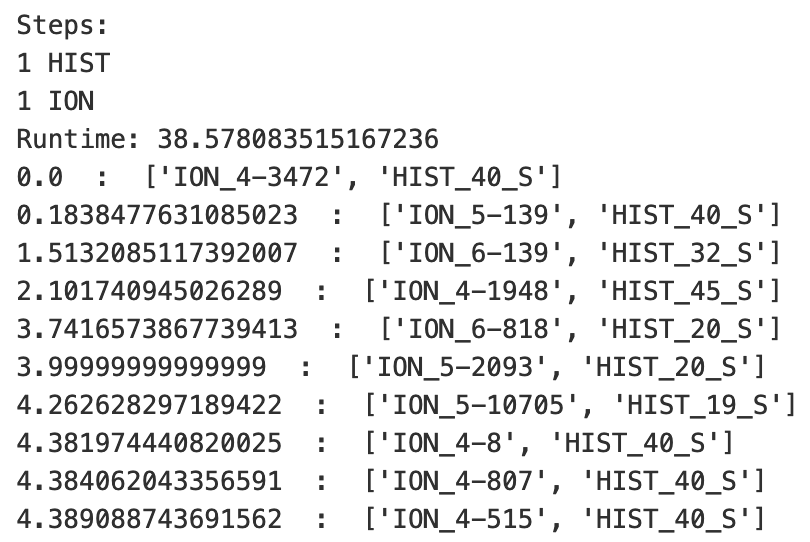}
}
\caption{Top ten results of the different runs.} 
\label{fig:run-results}
\end{figure}

\subsubsection{Date Range}
For date range, one run was completed including data over a period of three days, as can be seen in Table \ref{tab:range-running-times}, instead of including four weeks of measurements. However, this method took a long time compared to the others, because each individual date had to be filtered beforehand by checking if it is within the three day range. The additional computation time is not included in the run-time listed in the table. However, by reducing the date range to three days, we were able to reduce the step size and include more data points.
\subsection{Experimental Results}
For each run, the DTW calculates the distance for each ION and HIST pair and sorts them based on distance. The lower the distance, the more similar the data. The DTW distance results for the first 200 points stayed more stable in distance compared to the step size and the three day range results due to deviations in the data typically appearing later in the time series. The first 200 points and step size results have several of the same matches, while the top two results remain the same. The three day results are not as similar as the other two sample approaches, but they include some of the same top ten results. The results of the runs are shown in Figure \ref{fig:run-results}. Finally, we decided to focus on the step size results as an accurate similarity ranking due to its inclusion of the whole range of data and manageable run-time.

To validate the results of the DTW for step size, we look at the statistics for the top four matches and lowest-ranked four matches ranked by the DTW, seen in Table \ref{tab:statistics} and Table \ref{tab:statisticsWorst}.

The count for both the lowest and highest-ranked matches is similar with HIST possessing about 640 times more points compared to the ION measurements which is expected because HIST is collected at a higher frequency. While the standard deviation for the ION measurements for the top four matches is zero because they form a straight line, the standard deviation for the ION measurements for the lowest-ranked matches is higher. This implies that the ION measurements for the lowest-ranked matches are more spread out. 

Merging the ION and HIST data points for all the matches show that the four highest-ranked matches are more similar to each other compared to the lowest-ranked four matches. The standard deviation for the merged data for the top four matches stays constant, while the standard deviation for the lowest-ranked merged matches increases substantially compared to the individual standard deviations. This is because the HIST and ION measurements for the lowest-ranked four matches are less similar to each other and therefore the values will be more spread out when the data is merged which also leads to an increased data range. Based on these differences, we were able to verify that the DTW ranked the measurements correctly.

\begin{table}[!t]
\begin{center}
\begin{tabular}{ |p{1.7cm}||p{0.9cm}|p{0.9cm}|p{0.9cm}|p{0.9cm}|p{0.9cm}| }
 \hline
 measurement name & count & mean & std & min & max\\
 \hline
 ION-4-3472 & 744  & 0&  0 & 0 & 0\\
 HIST-40-S &  481302 & 0.0098  & 8.8 & -2048 & 1792\\
 Merged & 481892 & 0.0098 & 8.8 & -2048 & 1792\\
 \hline
 ION-5-139  & 739 & 0.013 & 0 & 0.013 & 0.013\\
 HIST-40-S & 481302 & 0.0098 & 8.8 & -2048 & 1792\\
 Merged & 481896 & 0.0098 & 8.8 & -2048 & 1792\\
 \hline
 ION-4-3472& 744 & 0 &0 & 0 &0\\
 HIST-44-S & 481301 & 0.014 & 9.4 & -2048 & 1748\\
 Merged & 481891 & 0.014 & 9.4 & -2048 & 1748\\
 \hline
 ION-5-139 & 739 & 0.013 & 0 & 0.013 & 0.013\\
 HIST-44-S & 481301 & 0.014 & 9.4 & -2048 & 1748\\
 Merged & 481895 & 0.014 & 9.4 & -2048 & 1748\\
 \hline
\end{tabular}
\end{center}
\caption{Statistics Top Four Ranked Matches.}
\label{tab:statistics}
\end{table}

\begin{table}[!t]
\begin{center}
\begin{tabular}{ |p{1.7cm}||p{0.9cm}|p{0.9cm}|p{0.9cm}|p{0.9cm}|p{0.9cm}| }
 \hline
 measurement name & count & mean & std & min & max\\
 \hline
 ION-4-198 & 756  & $6.5x10^7$&  188 & $6.5x10^7$ & $6.5x10^7$\\
 HIST-23-S &  891237 & -91.9  & 6.6 & -99 & 100\\
 Merged & 891730 & 54631 & 1878654 & -99 & $6.5x10^7$\\
 \hline
 ION-4-198 & 756  & $6.5x10^7$&  188 & $6.5x10^7$ & $6.5x10^7$\\
 HIST-17-S & 891237 & -565 & 46 & -782 & -465\\
 Merged & 891730 & 54159 & 1878668 & -782 & $6.5x10^7$\\
 \hline
 ION-4-198 & 756  & $6.5x10^7$&  188 & $6.5x10^7$ & $6.5x10^7$\\
 HIST-16-S & 891237 & -628 & 47& -785 & -538\\
 Merged & 891730 & 54096 & 1878670 & -785 & $6.5x10^7$\\
 \hline
 ION-4-198 & 756  & $6.5x10^7$&  188 & $6.5x10^7$ & $6.5x10^7$\\
 HIST-24-S & 891237 & -2048 & 0 & -2048 & -2048\\
 Merged & 891730 & 52677 & 1878711 & -2048 & $6.5x10^7$\\
 \hline
\end{tabular}
\end{center}
\caption{Statistics Four Lowest-Ranked Matches.}
\label{tab:statisticsWorst}
\end{table}
\subsection{Discussion}
The four most similar results from the DTW 100 and 2 step size are depicted in Figure \ref{fig: 4 results}. Most of the points for these graphs are concentrated on y = 0, with some HIST outlier points deviating from the axis. The ION measurements on the bottom left of each graph are a straight line while the HIST points on the bottom right have more spikes. The DTW algorithm matches them, because the HIST measurement points are concentrated on the same line as the ION counterpart, but with a few points that fall outside that line. The difference in pattern between the time series when looking at the bottom two graphs within each figure could mean one of two things, they could be different measurements; or they are the same but the ION measurements are not measured at a high enough frequency to record the spikes that appear in their HIST counterpart. The representative sample for the HIST measurements also does not include all the spikes that the actual measurements have because of the increased step size. This was the case with all the different methods of choosing a representative sample. Therefore, the HIST sample could have negatively affected the results of the DTW algorithm, because it does not include the entirety of spikes from the original data. The sample data is drawn in the top right corner for each graph in Figure \ref{fig: 4 results} and clearly shows less spikes compared to the actual HIST measurement.
After we obtained these results, we used the four most similar matches from the DTW 100 HIST, 2 ION steps for anomaly detection as illustrated in Figure \ref{fig: 4 results} and Figure \ref{fig:run-results}(a) (up to and including ION-5-139 and HIST-44-S).

\begin{table}
\begin{center}
\begin{tabular}{ |p{2cm}||p{1.7cm}|p{1.7cm}|p{1.7cm}| }
 \hline
 measurement name & Rolling Average & Autoregression & Level Shift\\
 \hline
 ION-4-3472 & 0  & 0&  0\\
 HIST-40-S &  94 & 276  &0\\
 Merged & 94 & 276 & 0\\
 \hline
 ION-5-139  & 0 & 0 & 0\\
 HIST-40-S &  94 & 276 & 0\\
 Merged & 832 & 274  & 4\\
 \hline
 ION-4-3472& 0 & 0 &0\\
 HIST-44-S & 221 & 804 & 6\\
 Merged & 221 & 804 & 4\\
 \hline
 ION-5-139 & 0 & 0 & 0\\
 HIST-44-S & 221 & 804 & 6\\
 Merged & 959 & 2269 & 16\\
 \hline
\end{tabular}
\end{center}
\caption{Number of Anomalies Top Four Matches.}
\label{tab:anomalies}
\end{table}

\section{Anomaly Detection}
\label{sec:AD}
\subsection{Implementation}
We took the top four matches of the DTW results and applied autoregression (AR), level shift (LS), and rolling average (RA) on them. Each anomaly detection algorithm is used separately on each matched pair. The data of the matched pairs are then combined into one time series to provide the same context and to possibly reduce the run-time of anomaly detection by running the algorithms on one time series compared to two. After combining the data, anomaly detection is applied on the time series as seen in Table \ref{tab:anomalies}.

\begin{figure*}[!t]
  \centering
  \includegraphics[width=.9\textwidth]{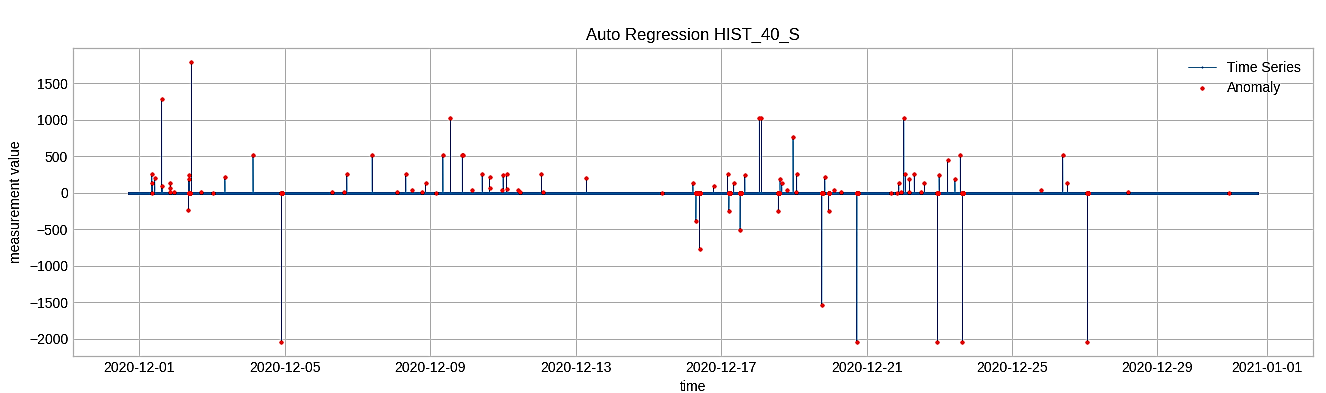}
  \caption{Autoregression HIST 40 S}
  \label{fig:AR_HIST}
\end{figure*}

\begin{figure*}[!t]
  \centering
  \includegraphics[width=.9\textwidth]{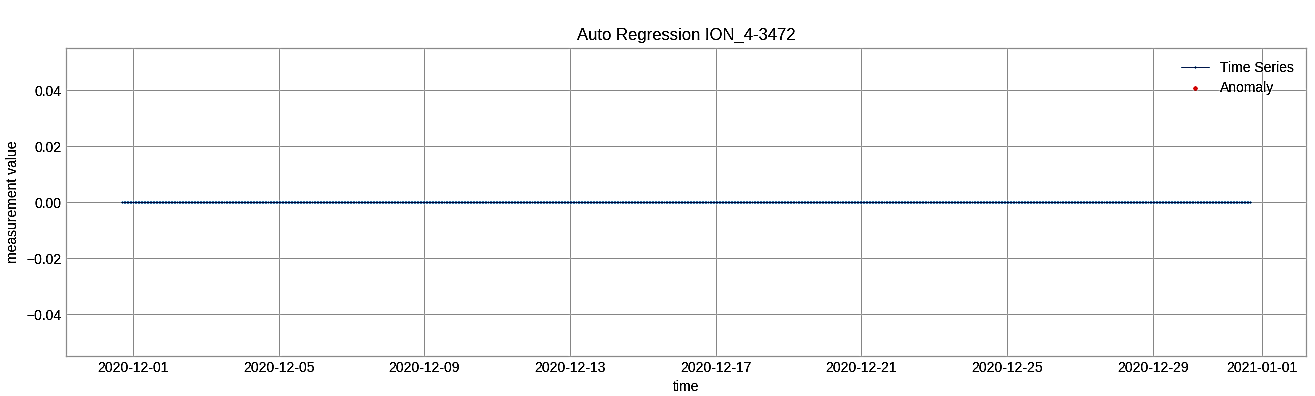}
  \caption{Autoregression ION 4 3472}
  \label{fig:AR_ION}
\end{figure*}
\subsection{Results}
For the top four matches, up to 94 to 804 times more anomalies were found when considering anomalies detected by both overlapping measurements compared to looking at a single measurement. These numbers depend on the anomaly detection algorithm applied on the data. Most anomalies are missed when only the ION measurements are considered instead of both HIST and ION measurements. When merging the data of the top four matches, a percent change of up to 182 to 785\% in anomalies can be detected compared to not merging the data. Merging the data therefore provides a more holistic view, because more anomalies are detected. However, identifying critical anomalies and false positives from the detected anomalies is a challenge because the data set is unlabeled. 
When looking at the number of anomalies, all ION measurements have no anomalies because they are a straight line or are close to forming a straight line. For the top four anomaly matches, autoregression discovers the most anomalies after rolling average, and then level shift. Level shift is less susceptible to noisy data and therefore detects the least number of anomalies. When the ION and HIST measurements were combined for the top four matches, the amount of anomalies either stayed the same or increased except for ION-4-3472 and HIST-44-S during level shift anomaly detection. This supports the point that merging the measurements leads to a more holistic view. The differences in anomalies among DTW matches are displayed in Figure \ref{fig:AR_HIST} and \ref{fig:AR_ION}. Figure \ref{fig:AR_HIST} has anomalies mostly along its spikes. In comparison, Figure \ref{fig:AR_ION} is a straight line with no deviations and therefore has no anomalies. 

\subsection{Discussion}
\label{sec:discussion}

Since autoregression works by comparing its previous value to the current one, the spikes should appear one at a time which also highlights that the step size sample approach might have easily missed the spikes because they are relatively rare. However, the mismatch in anomalies highlights the need to look at both HIST and ION measurements instead of only looking at one of them. The spikes, which were identified as anomalies, would have been missed by just looking at the ION measurements. Additionally, using both HIST and ION for anomaly detection increases the amount of data points and decreases the time between measurements. Therefore, using both measurements can increase the accuracy of anomaly detection, because they provide a more holistic view of the same measurement. The challenge with evaluating the accuracy of the different anomaly detection algorithms is that the data is unlabeled and it is unknown if the anomalies that are identified by the anomaly detection algorithms are cause for concern (i.e., "critical"). Critical anomalies can help with detecting cyber attacks or unusual activity in the system, while naturally occurring anomalies are common to the system. Further research needs to be done on how to evaluate the anomaly detection models and how to identify critical anomalies and false positives.

\section{Conclusion}
\label{sec:conc}
Overall, this paper examines approaches used with overlapping electrical measurements from a power grid with the goal of improving anomaly detection. While there were challenges with the evaluation of the unsupervised anomaly detection, it was demonstrated that merging the overlapping measurements extends its time frame and provides a more holistic view of the data. In general, this research identified ways to incorporate overlapping measurements for unsupervised anomaly detection. The DTW algorithm matched the most similar measurements, and we were able to validate the matching results by looking at the graphs and statistics of the highest-ranked and lowest-ranked matches. By using DTW, we were able to identify more anomalies. The number of anomalies identified by the anomaly detection algorithms differ among the overlapping measurement matches which provide additional challenges that need to be addressed. The difference in frequency between the two independent system introduces challenges that are most likely impacting the DTW results and the difference in anomalies. Overall, this work is important, because anomaly detection mechanism can serve as an early warning system when trying to detect cyber attacks.

\section{Future Work}
\label{sec:fw}
Future work includes analyzing other statistical methods that assess when the differences in anomalies are relevant enough that the measurements are not the same. Especially, investigating the mismatch between anomalies for the same measurements, which could indicate certain types of data tampering. In addition, exploring different ways to filter the data before running DTW to reduce run-times and improve the scalability of the approach. Using the pearson correlation coefficient would be a possible option to find similar time series within each system and then filter based on the results.

Another possible area for future work includes determining which anomaly detection algorithm is the most accurate one. A possible approach would be to simulate an artificial anomaly and see which algorithm detects it. If the zero measurements are excluded from the overall data, a Denial-of-service (DoS) attack can be inserted. A DoS attack is one of the most common threats to synchrophasor systems \cite{article} and overwhelms a PMU with bogus frames so that legitimate frames are lost, delayed, denied, or dropped \cite{8982275}. The consequences include that the real-time measurements would be delayed or dropped. For this case, we considered inserting zero measurements over a time period of about six to eight seconds to indicate a DoS attack. The zero measurements indicate dropped packages during transmission and would show up when looking at the database of sensor measurements. Ideally, both overlapping measurements would more accurately identify the anomalies and allow us to compare the accuracy of the different anomaly detection algorithms. 

The same could be done for a data integrity attack, which includes tampering with the signal measurements units of devices through interference, or changing calibration, forging data, or even GPS spoofing \cite{8982275}. We considered adding artificial anomalies, for example, adding noisy data points from the Gaussian distribution. Because the HIST measurements are already noisy due to the many spikes, adding the artificial points would make it difficult to distinguish between the actual attack and the occurring spikes. The cause for the spikes in the data needs to be defined first to be able to include a data integrity attack. If the spikes are naturally occurring noisy data points due to poor calibration, these spikes could be removed and then a data integrity attack can be inserted.  

Evaluating our approach using overlapping measurements on a labeled data set to confirm the effectiveness is another direction of future work. One possible approach is generating a synthetic data set and labeling it to change it to a supervised anomaly detection problem which has more known methods. However, the shape of the data needs to be explained, in addition to defining the anomalies in more detail to make sense with the shape.

\ifCLASSOPTIONcompsoc
 \section*{Acknowledgments}
\else
 \section*{Acknowledgment}
\fi

The authors would like to thank the Information Science and Technology Institute (ISTI) at Los Alamos National Laboratory. This work is supported in part by the National Science Foundation under Grant 2043324 and 2025682.



%

\bibliographystyle{IEEEtran}
\bibliography{IEEEabrv,cps}

\begin{thebibliography}{10}
\providecommand{\url}[1]{#1}
\csname url@samestyle\endcsname
\providecommand{\newblock}{\relax}
\providecommand{\bibinfo}[2]{#2}
\providecommand{\BIBentrySTDinterwordspacing}{\spaceskip=0pt\relax}
\providecommand{\BIBentryALTinterwordstretchfactor}{4}
\providecommand{\BIBentryALTinterwordspacing}{\spaceskip=\fontdimen2\font plus
\BIBentryALTinterwordstretchfactor\fontdimen3\font minus
  \fontdimen4\font\relax}
\providecommand{\BIBforeignlanguage}[2]{{%
\expandafter\ifx\csname l@#1\endcsname\relax
\typeout{** WARNING: IEEEtran.bst: No hyphenation pattern has been}%
\typeout{** loaded for the language `#1'. Using the pattern for}%
\typeout{** the default language instead.}%
\else
\language=\csname l@#1\endcsname
\fi
#2}}
\providecommand{\BIBdecl}{\relax}
\BIBdecl

\bibitem{sontowski2020cyber}
S.~Sontowski, M.~Gupta, S.~S.~L. Chukkapalli, M.~Abdelsalam, S.~Mittal,
  A.~Joshi, and R.~Sandhu, ``Cyber attacks on smart farming infrastructure,''
  in \emph{2020 IEEE 6th International Conference on Collaboration and Internet
  Computing (CIC)}.\hskip 1em plus 0.5em minus 0.4em\relax IEEE, 2020, pp.
  135--143.

\bibitem{hahn2011cyber}
A.~Hahn and M.~Govindarasu, ``Cyber attack exposure evaluation framework for
  the smart grid,'' \emph{IEEE Transactions on Smart Grid}, vol.~2, no.~4, pp.
  835--843, 2011.

\bibitem{gupta2018authorization}
M.~Gupta and R.~Sandhu, ``Authorization framework for secure cloud assisted
  connected cars and vehicular internet of things,'' in \emph{Proceedings of
  the 23nd ACM on symposium on access control models and technologies}, 2018,
  pp. 193--204.

\bibitem{le2020gridattacksim}
T.~D. Le, A.~Anwar, S.~W. Loke, R.~Beuran, and Y.~Tan, ``Gridattacksim: A cyber
  attack simulation framework for smart grids,'' \emph{Electronics}, vol.~9,
  no.~8, p. 1218, 2020.

\bibitem{zhu2011taxonomy}
B.~Zhu, A.~Joseph, and S.~Sastry, ``A taxonomy of cyber attacks on scada
  systems,'' in \emph{2011 International conference on internet of things and
  4th international conference on cyber, physical and social computing}.\hskip
  1em plus 0.5em minus 0.4em\relax IEEE, 2011, pp. 380--388.

\bibitem{zhou2016abnormal}
Y.~Zhou, R.~Arghandeh, I.~Konstantakopoulos, S.~Abdullah, A.~von Meier, and
  C.~J. Spanos, ``Abnormal event detection with high resolution micro-pmu
  data,'' in \emph{2016 Power Systems Computation Conference (PSCC)}.\hskip 1em
  plus 0.5em minus 0.4em\relax IEEE, 2016, pp. 1--7.

\bibitem{deka2015optimal}
D.~Deka, R.~Baldick, and S.~Vishwanath, ``Optimal data attacks on power grids:
  Leveraging detection \& measurement jamming,'' in \emph{2015 IEEE
  International Conference on Smart Grid Communications (SmartGridComm)}.\hskip
  1em plus 0.5em minus 0.4em\relax IEEE, 2015, pp. 392--397.

\bibitem{8762213}
V.~Venkataramanan, A.~K. Srivastava, A.~Hahn, and S.~Zonouz, ``Measuring and
  enhancing microgrid resiliency against cyber threats,'' \emph{IEEE
  Transactions on Industry Applications}, vol.~55, no.~6, pp. 6303--6312, 2019.

\bibitem{6003810}
C.-W. Ten, J.~Hong, and C.-C. Liu, ``Anomaly detection for cybersecurity of the
  substations,'' \emph{IEEE Transactions on Smart Grid}, vol.~2, no.~4, pp.
  865--873, 2011.

\bibitem{chandola2009anomaly}
V.~Chandola, A.~Banerjee, and V.~Kumar, ``Anomaly detection: A survey,''
  \emph{ACM computing surveys (CSUR)}, vol.~41, no.~3, pp. 1--58, 2009.

\bibitem{moghaddass2017hierarchical}
R.~Moghaddass and J.~Wang, ``A hierarchical framework for smart grid anomaly
  detection using large-scale smart meter data,'' \emph{IEEE Transactions on
  Smart Grid}, vol.~9, no.~6, pp. 5820--5830, 2017.

\bibitem{gupta2021detecting}
D.~Gupta, M.~Gupta, S.~Bhatt, and A.~S. Tosun, ``Detecting anomalous user
  behavior in remote patient monitoring,'' in \emph{IEEE IRI}, 2021.

\bibitem{fenza2019drift}
G.~Fenza, M.~Gallo, and V.~Loia, ``Drift-aware methodology for anomaly
  detection in smart grid,'' \emph{IEEE Access}, vol.~7, pp. 9645--9657, 2019.

\bibitem{cook2019anomaly}
A.~A. Cook, G.~M{\i}s{\i}rl{\i}, and Z.~Fan, ``Anomaly detection for iot
  time-series data: A survey,'' \emph{IEEE Internet of Things Journal}, vol.~7,
  no.~7, pp. 6481--6494, 2019.

\bibitem{adkisson2021autoencoder}
M.~Adkisson, J.~C. Kimmel, M.~Gupta, and M.~Abdelsalam, ``Autoencoder-based
  anomaly detection in smart farming ecosystem,'' \emph{arXiv preprint
  arXiv:2111.00099}, 2021.

\bibitem{gupta2021hierarchical}
D.~Gupta, O.~Kayode, S.~Bhatt, M.~Gupta, and A.~S. Tosun, ``Hierarchical
  federated learning based anomaly detection using digital twins for smart
  healthcare,'' \emph{arXiv preprint arXiv:2111.12241}, 2021.

\bibitem{9096053}
A.~Barua, D.~Muthirayan, P.~P. Khargonekar, and M.~A. Al~Faruque,
  ``Hierarchical temporal memory based machine learning for real-time,
  unsupervised anomaly detection in smart grid: Wip abstract,'' in \emph{2020
  ACM/IEEE 11th International Conference on Cyber-Physical Systems (ICCPS)},
  2020, pp. 188--189.

\bibitem{6877810}
G.~Elafoudi, L.~Stankovic, and V.~Stankovic, ``Power disaggregation of domestic
  smart meter readings using dynamic time warping,'' in \emph{2014 6th
  International Symposium on Communications, Control and Signal Processing
  (ISCCSP)}, 2014, pp. 36--39.

\bibitem{9299915}
J.~R. Ausmus, P.~K.~P. Sen, T.~Wu, U.~Adhikari, Y.~Zhang, and V.~Krishnan,
  ``Improving the accuracy of clustering electric utility net load data using
  dynamic time warping,'' in \emph{2020 IEEE/PES Transmission and Distribution
  Conference and Exposition (T D)}, 2020, pp. 1--5.

\bibitem{8919604}
D.~M. Diab, B.~AsSadhan, H.~Binsalleeh, S.~Lambotharan, K.~G. Kyriakopoulos,
  and I.~Ghafir, ``Anomaly detection using dynamic time warping,'' in
  \emph{2019 IEEE International Conference on Computational Science and
  Engineering (CSE) and IEEE International Conference on Embedded and
  Ubiquitous Computing (EUC)}, 2019, pp. 193--198.

\bibitem{9178465}
Z.~Zheng, M.~Zhou, Y.~Chen, M.~Huo, L.~Sun, S.~Zhao, and D.~Chen, ``A fused
  method of machine learning and dynamic time warping for road anomalies
  detection,'' \emph{IEEE Transactions on Intelligent Transportation Systems},
  pp. 1--13, 2020.

\bibitem{AutoregressionNetwork}
Y.~Zhou and J.~Li, ``Research of network traffic anomaly detection model based
  on multilevel autoregression,'' in \emph{2019 IEEE 7th International
  Conference on Computer Science and Network Technology (ICCSNT)}, 2019, pp.
  380--384.

\bibitem{ARIMACloud}
F.~Schmidt, F.~Suri-Payer, A.~Gulenko, M.~Wallschläger, A.~Acker, and O.~Kao,
  ``Unsupervised anomaly event detection for cloud monitoring using online
  arima,'' in \emph{2018 IEEE/ACM International Conference on Utility and Cloud
  Computing Companion (UCC Companion)}, 2018, pp. 71--76.

\bibitem{ARIMANetwork}
A.~H. Yaacob, I.~K. Tan, S.~F. Chien, and H.~K. Tan, ``Arima based network
  anomaly detection,'' in \emph{2010 Second International Conference on
  Communication Software and Networks}, 2010, pp. 205--209.

\bibitem{ChangePointRadar}
O.~Solomentsev, M.~Zaliskyi, and T.~Gerasymenko, ``Change-point detection
  during radar operation,'' in \emph{2016 IEEE First International Conference
  on Data Stream Mining Processing (DSMP)}, 2016, pp. 295--298.

\bibitem{ChangePointSensor}
J.~Geng and L.~Lai, ``Bayesian quickest change point detection and localization
  in sensor networks,'' in \emph{2013 IEEE Global Conference on Signal and
  Information Processing}, 2013, pp. 871--874.

\bibitem{DTWMueen}
\BIBentryALTinterwordspacing
A.~Mueen and E.~Keogh, ``Extracting optimal performance from dynamic time
  warping,'' in \emph{Proceedings of the 22nd ACM SIGKDD International
  Conference on Knowledge Discovery and Data Mining}, ser. KDD '16.\hskip 1em
  plus 0.5em minus 0.4em\relax New York, NY, USA: Association for Computing
  Machinery, 2016, p. 2129–2130. [Online]. Available:
  \url{https://doi.org/10.1145/2939672.2945383}
\BIBentrySTDinterwordspacing

\bibitem{salvador2004fastdtw}
S.~Salvador and P.~Chan, ``Fastdtw: Toward accurate dynamic time warping in
  linear time and space,'' in \emph{KDD workshop on mining temporal and
  sequential data}.\hskip 1em plus 0.5em minus 0.4em\relax Citeseer, 2004.

\bibitem{9166366}
E.~H. Budiarto, A.~Erna~Permanasari, and S.~Fauziati, ``Unsupervised anomaly
  detection using k-means, local outlier factor and one class svm,'' in
  \emph{2019 5th International Conference on Science and Technology (ICST)},
  vol.~1, 2019, pp. 1--5.

\bibitem{medico2020}
R.~Medico, ``rob-med/awesome-ts-anomaly-detection,'' Aug 2020.

\bibitem{hannon2021real}
C.~Hannon, D.~Deka, D.~Jin, M.~Vuffray, and A.~Y. Lokhov, ``Real-time anomaly
  detection and classification in streaming pmu data,'' in \emph{2021 IEEE
  Madrid PowerTech}.\hskip 1em plus 0.5em minus 0.4em\relax IEEE, 2021, pp.
  1--6.

\bibitem{ratanamahatana2005}
C.~Ratanamahatana and E.~Keogh, ``Three myths about dynamic time warping data
  mining,'' 04 2005.

\bibitem{article}
S.~Grewal, M.~Soni, and D.~Jain, ``Cyber security threats in synchrophasor
  system in wide area monitoring system,'' \emph{TELKOMNIKA Indonesian Journal
  of Electrical Engineering}, vol.~15, pp. 436--444, 09 2015.

\bibitem{8982275}
A.~Sundararajan, T.~Khan, A.~Moghadasi, and A.~I. Sarwat, ``Survey on
  synchrophasor data quality and cybersecurity challenges, and evaluation of
  their interdependencies,'' \emph{Journal of Modern Power Systems and Clean
  Energy}, vol.~7, no.~3, pp. 449--467, 2019.

\end{thebibliography}

\end{document}